# Does the Universe Have a Handedness?

Michael J. Longo

Department of Physics, University of Michigan, Ann Arbor, MI 48109

In this article I extend an earlier study of spiral galaxies in the Sloan Digital Sky Survey (SDSS) to investigate whether the universe has an overall handedness. A preference for spiral galaxies in one sector of the sky to be left-handed or right-handed spirals would indicate a parity-violating asymmetry in the overall universe and a preferred axis. The previous study used 2616 spiral galaxies with redshifts <0.04 and identified handedness. The new study uses 15872 with redshifts <0.085 and obtains very similar results to the first with a signal exceeding 5 $\sigma$, corresponding to a probability of $2.5 \times 10^{-7}$ for occurring by chance. The axis of the dipole asymmetry lies at approx. ($l, b$) =(32°,69°), roughly along that of our Galaxy and close to the so-called "Axis of Evil".

**PACS:** 98.65.-r, 98.62.Py, 98.80.-k, 98.52.Eh

### 1. INTRODUCTION

Symmetry has a strong appeal to the human psyche. Nature, however, exhibits some surprising asymmetries. On the smallest scales, an asymmetry (parity violation) was found in the angular distribution of electrons in the beta decay of spin oriented $^{60}$Co, confirming the proposal by Lee and Yang that parity was violated in weak decays[1]. On the molecular scale, there is a large predominance of left-handed amino acids over right-handed ones in organisms, the origin of which is still not well understood. It is reasonable to ask if nature exhibits such an asymmetry on the largest scales.

Spiral galaxies with a well-defined handedness offer a means to test this possibility. Ideally the signal for such an asymmetry would be an excess of one handedness in a large region of the sky and a similar excess of the opposite handedness in the opposite direction (i.e., a dipole). The preponderance of data in the northern Galactic hemisphere, as well as the masking of much of the sky by dust in the Milky Way, complicates the search for such an effect. However, the spiral handedness technique has important advantages in that it is not biased by the incompleteness of the maps or by atmospheric or instrumental effects, which cannot turn right-handed spirals into left-handed. One has to be careful of an overall bias due to a preference toward assigning left-handed or right-handed. Such a bias would show up as a "monopole". In principle, a computer algorithm to assign handednesses could be developed[2,3], but in practice the huge range of galaxy brightness, color, size on the sky, orientation, and structure makes this exceedingly difficult to do efficiently[a]. In this study, as in the Galaxy Zoo study[4], human scanners chose the spiral sample and made the handedness assignments. Precautions against such a left/right bias will be discussed below.

In the first study[2,3] I used galaxies from the SDSS DR5 database[5] that contains ~40,000 galaxies with spectra for redshifts <0.04. In this study I use the DR6 database[6] with ~230,000 galaxies with redshifts <0.085. A few percent of these are spiral galaxies with identifiable handedness that can be used in the study.

## 2. THE ANALYSIS

Objects classified as "galaxies" in the SDSS DR6 database were used in this analysis. A list of galaxies that had measured redshifts less than 0.10 was obtained from the SDSS DR6 web sites, cas.sdss.org/dr6 and casjobs.sdss.org. Spiral galaxies are typically bluer than elliptical ones. Strateva et al.[7] show that elliptical galaxies can be separated from spirals fairly cleanly by a $z$-dependent cut on $(u-r)$ where $u$ and $r$ are the apparent magnitudes for the ultraviolet (354 nm) and red (628 nm) bands respectively[8]. A conservative cut to enhance the fraction of spirals was therefore made by requiring that $(u-r) < 2.85$. Galaxy images from the resulting list of ~200,000 galaxy images were then looked at by a team of 5 scanners.

Individual RGB images of the galaxies from the list were acquired from the SDSS web site and displayed to the scanners using an HTML/JAVA program. The HTML program mirrored half of the images at random to reduce scanning biases favoring a particular handedness.[b] The scanners had no visual cue as to whether the image was mirrored. Scanners were assigned small $z$ slices at random, and the scanning was done in random order with respect to right ascension, $\alpha$, and declination $\delta$ so that any scanning bias could not cause a systematic bias on the $(z, \alpha, \delta)$ distributions of the handedness, only a possible overall bias in the complete sample. The scanners had only 3 choices: <u>Left</u>, <u>Right</u>, or <u>Unclear</u>, where $Left \equiv \circlearrowleft$ and $Right \equiv \circlearrowright$. No attempt to otherwise classify the galaxies was made. Scanners were instructed to classify galaxies as Unclear unless the handedness was clear. Overall, about 15% of the galaxies were classified as having recognizable handedness ($L$ or $R$). No further analysis of the $U$'s was done.

---

[a] A program using a rotating mask is described in Version 1 of Ref. 2. It gave similar results to the human scanning used in later versions, but was considerably less efficient, especially at larger redshifts.

[b] Ideally, to eliminate biases completely, the complete lists should be scanned twice, intermingling all the images with their mirrored versions, but this would have taken another year of work. No record of the mirroring was kept.

To reproduce the earlier study[2, 3] as closely as possible, I required that the green magnitude be <17 for redshifts $z < 0.04$. Beyond $z=0.04$ the magnitude limit was increased to 17.4 at a redshift of 0.10. The handedness of galaxies fainter than magnitude 17.4 and galaxies with $z > 0.085$ were generally difficult to classify, so these were not used.

As in Refs. 2, 3, a cut was made to remove the bluest galaxies that tend to be those with recent star formation initiated by a collision. This required $(u-z) > 1.6$, where $z$ is the apparent magnitude in the far infrared; it removed 1.9% of the $L+R$ sample. A cut to remove the reddest galaxies, $(u-z) < 3.5$ was also made; this removed 2.7% of the sample, leaving 15,158.

Most of the SDSS DR6 data is in the northern Galactic hemisphere ($\alpha \sim 192°$, $\delta \sim 27°$) with declinations $-5° < \delta < 63°$. In the southern Galactic hemisphere there is only coverage in 3 narrow bands in $\delta$ near $\delta = -10°, 0°$, and $14°$, each about 4° wide. In Refs. 2, 3 I used all declinations between -19° and +60° which includes most of the data. In this analysis I use the same declination ranges. This left 14683 galaxies compared to 2616 in the first study.

None of these cuts or the incomplete coverage of the survey would be expected to cause a bias between left- and right-handed spirals.

## 3. RESULTS

A plot of asymmetries $\langle A \rangle \equiv (R-L)/(R+L)$, binned in sectors of right ascension and slices in $z$, is shown in Fig. 1. Positive $\langle A \rangle$ are shown in red and negative ones in blue. The larger numbers near the perimeter give the net asymmetry for the entire right ascension sector. The black numbers in parentheses next to them give the total number of galaxies in that sector. The $\sigma$ are determined from standard normal distribution statistics, $\sigma(N) = \sqrt{N}$, which gives $\sigma(\langle A \rangle) = 1/\sqrt{R+L}$. There is an apparent excess of left-handed spirals in the sectors for $150° < \alpha < 240°$. There is a complementary excess of right-handed in the opposite hemisphere, though there are only 1/7$^{th}$ as many galaxies there.

The incompleteness of the survey, especially in $\delta$, makes a complete multipole analysis of the asymmetry data of dubious value. In any case a preferred spiral handedness implies a dipole component and the lack of a monopole (bias), so I restrict my analysis to these two terms.

*Bias*–The galaxies were scanned in random order with respect to ($\alpha$, $\delta$, $z$), so no position dependent bias is possible. Also, half the galaxies were randomly mirrored during scanning and precautions against left-right bias were taken in the web interface used by the scanners. The best check for an overall bias is to look at the complete scanned sample that included galaxies at larger $z$ and fainter luminosities as well as some that were scanned more than once. This sample included 25612 galaxies and gave $R=12707$, $L=12905$ and an overall asymmetry of $-0.0077 \pm 0.0062$, even though this sample included 7 times as many galaxies in the $150° < \alpha < 240°$ sector with its apparent excess of $L$ spirals. If the $150°<\alpha<240°$ sector is removed, the asymmetry becomes $+0.0133\pm0.010$, consistent with no bias or a small positive one.

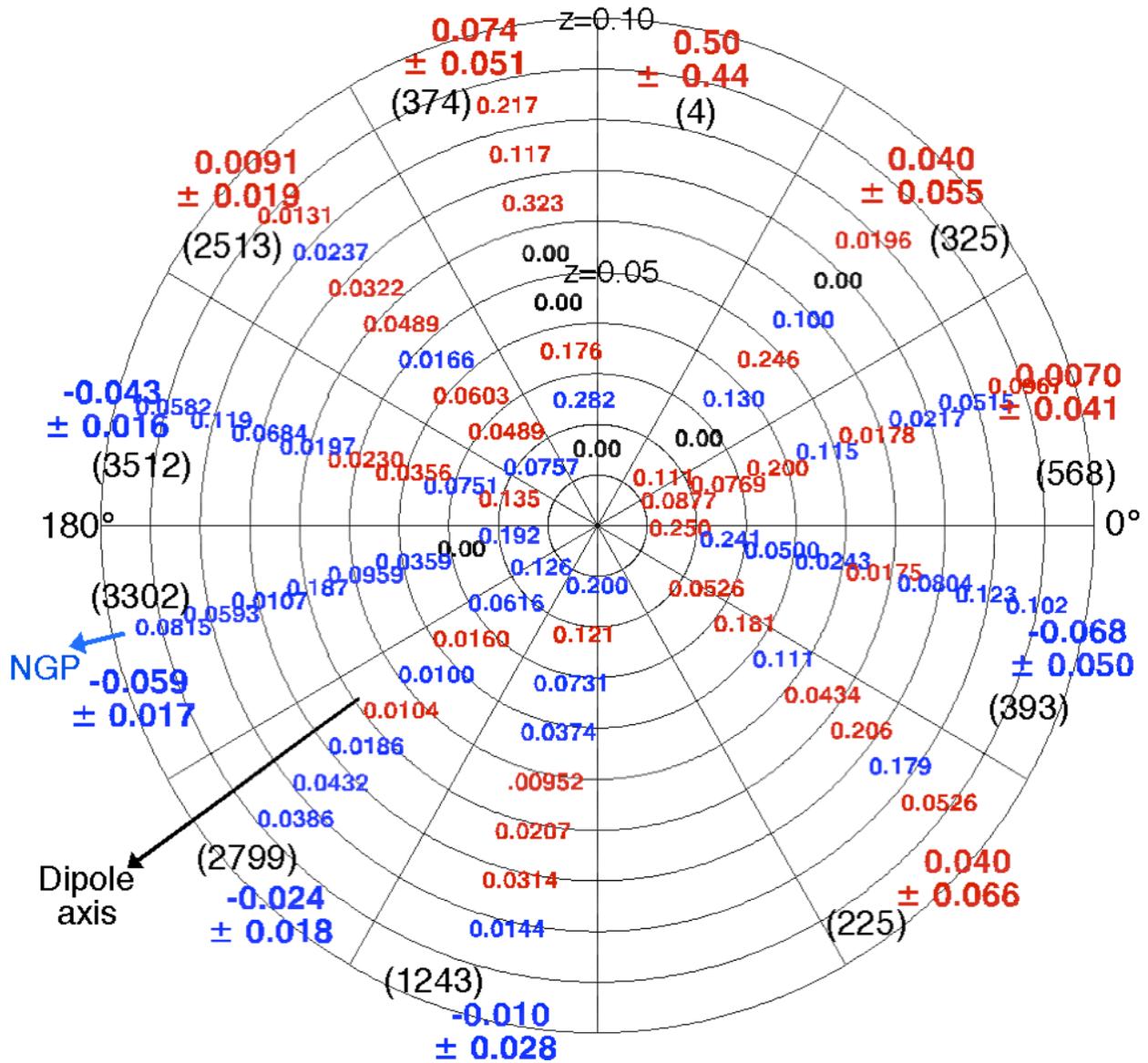

FIG. 1. Polar plot of net asymmetries $\langle A \rangle$ by sector in right ascension and segments in $z$. Segments with positive $\langle A \rangle$ are indicated in red and negative $\langle A \rangle$ in blue. The $\langle A \rangle$ for segments with <10 galaxies are not shown. The larger numbers near the periphery give the overall asymmetry for that sector; the black numbers in parentheses are the total number of spiral galaxies in the sector. The NGP is the north pole of our Galaxy. The black arrow shows the most probable dipole axis. Declinations between -19° and +60° were used.

*The Dipole*–To investigate a dipole, the asymmetries were binned using the 12 $\alpha$ sectors in Figure 1 and 4 bins in $\delta$. This gave a total of 31 bins with nonzero contents. Then a particular right ascension and declination was chosen as an axis. The space angle $\gamma_i$ between this axis and the average $(\alpha, \delta)$ of the binned data was then calculated and the asymmetries were binned in $\gamma_i$. This gave 31 averaged asymmetries $A(\gamma_i)$. The $A(\gamma_i)$ were fitted to an $a_0 \cos\gamma_i$ dependence. A grid of axis $(\alpha, \delta)$ was then searched to find the axis that gave the minimum $\chi^2$/dof. This occurred at $(\alpha, \delta) = (216°, 25°)$, or $(l, b) = (32°, 69°)$ in Galactic coordinates. The uncertainty in space angle is $\sim 17°$. The $a_0 \cos\gamma$ fit gave a $\chi^2$/dof $= 0.819$ and a dipole term $a_0 = -0.0402\pm0.0109$. Thus the data are consistent with a pure dipole with a significance of $3.7\sigma$, corresponding to a probability of $2.2\times10^{-4}$ for the dipole to be equal 0. A plot of the asymmetry *vs.* $\cos\gamma$ is shown in Figure 2 along with the best fit to a pure dipole. If a constant term (monopole or bias) $a_1$ is allowed, then the $\chi^2$/dof <u>increases</u> to 0.842 with $a_1 = -0.0050\pm0.0063$.

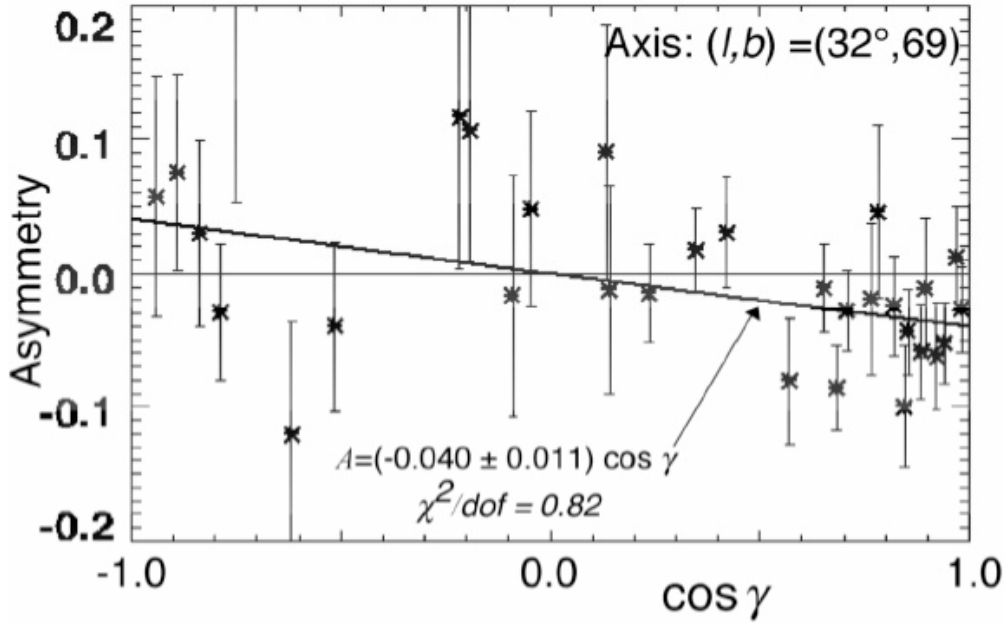

FIG. 2. Asymmetry vs. $\cos\gamma$ at the optimal dipole axis for the 31 $(\alpha, \delta)$ bins. The straight line is a fit to an $a_0 \cos\gamma_i$ dependence.

*Statistical Significance*–In Refs. 2, 3, in order to determine the overall statistical significance of the apparent asymmetry I used data in the declination range $-19°< \delta< 60°$, and the right ascension ranges $195°\pm45°$ and $0°\pm40°$. Here I use the same $\delta$ range and right ascensions in a narrower range $195°\pm30°$ and $15°\pm30°$. Table I shows the resulting asymmetries and their uncer-tainties. The 165°<$\alpha$<225° sector with 86% of the galaxies shows an asymmetry of −0.0695± 0.0127, a 5.48$\sigma$ effect. The data in the sparsely covered southern Galactic hemisphere (−15°<$\alpha$<45°) show a small positive asymmetry as would be expected for a real signal. Overall the asymmetry is -0.0607± 0.0118, a 5.15$\sigma$ effect with a probability of 2.5x10$^{-7}$ for occurring by chance.

Table I. Number counts and net asymmetries $\langle A \rangle$=($N_R$− $N_L$)/$N_{Tot}$ for the right ascension ranges indicated. The last two columns give the number of standard deviations for the $\langle A \rangle$ and the probability.

| $\alpha$ Range | $N_R$ | $N_L$ | $N_{Tot}$ | $\langle A \rangle \pm \sigma$ | $\langle A \rangle / \sigma$ | Prob. |
|---|---|---|---|---|---|---|
| -15° to 45° | 495 | 490 | 985 | 0.005±0.032 | +0.16 | 0.87 |
| 165° to 225° | 2890 | 3322 | 6212 | −0.0695±0.0127 | −5.48 | 2.1x10$^{-8}$ |
| Overall | | | 7197 | -0.0607±0.0118 | −5.15 | 2.5x10$^{-7}$ |

## 4. COMPARISON WITH OTHER STUDIES

*Iye and Sugai*–Iye and Sugai[9] have published a catalog of spin orientations of galaxies in the southern Galactic hemisphere that contains 8287 spiral galaxies. Of these, 3118 had *R* or *L* handedness about which both scanners agreed. I have analyzed their catalog using a sector opposite 165°<$\alpha$<225°, -5°< $\delta$ <+60°; i.e., −40° <$\alpha$<+45° and -60°< $\delta$ <+5°. Redshifts of most of the galaxies were not measured, so only their ($\alpha, \delta$) were used. This resulted in an asymmetry |*A*| = 0.047±0.029 with a preponderance of right-handed spirals in the southern Galactic hemisphere, in excellent agreement with the asymmetry 0.0607 ± 0.0118 that I observe with a preponderance of left-handed spirals in the northern hemisphere. It provides an independent confirmation of an asymmetry at the 1.6$\sigma$ level. Without ($\alpha, \delta$) cuts their overall asymmetry was 0.000±0.014, consistent with no bias in their study.

*Galaxy Zoo*–Galaxy Zoo[10] is an online project in which >100,000 volunteers visually classify the morphologies of galaxies selected from the spectroscopic sample of the SDSS DR6, the same sample used here. In K. Land et al. (Ref. 4) they investigate the possibility of a large scale spin anisotropy. Each galaxy was classified an average of 39 times. Those galaxies for which over 80% of the votes agreed constituted their "clean" sample and over 95% their "superclean" sample. They found that there was a large *L/R* bias in their samples. Their clean sample contained 17100 *L* (clockwise) spirals and 18471 *R*; the superclean sample contained 6106 *L* and 7034 *R*

spirals[c]. This gives an asymmetry (bias) of –0.0386±0.0053 for the clean sample and a much larger bias of –0.0706±0.0087 for the superclean sample that presumably contained more clearly recognizable spirals. This should be compared to the upper limit of –0.0077 ± 0.0062 found in this study as discussed in the section on biases above. Land et al. attributed these biases to the design of the Galaxy Zoo website or to a human pattern recognition effect that was shared by all their volunteers. A later bias study, mainly with the superclean sample, that compared monochrome images with mirrored RGB images found similar biases. They corrected for the bias by requiring only 78% agreement between scanners for the $R$ galaxies in the clean sample and 94% in the superclean. They assumed the bias was independent of redshift and magnitude, despite the fact that it is much easier to correctly assign the handedness of larger, brighter galaxies. In their analysis they found a dipole term of about $2\sigma$ along $(\alpha, \delta)=(161°,11°)$ consistent with the axis I found in Ref. 2 at (202°,25°). When a monopole (bias) term was also allowed this became a $1\sigma$ effect.

It is difficult to compare this study with the Galaxy Zoo result. They used galaxies with redshifts out to 0.3, whereas I used those with $z$<0.085 and restricted the magnitude range because of the difficulty in assigning the handedness of fainter galaxies. Their large biases also caused large uncertainties in the monopole/bias term, while the biases in this analysis were consistent with 0.

## 5. DISCUSSION AND CONCLUSIONS

The data show a strong signal for a preferred axis for redshifts <0.085. Overall the asymmetry is -0.0607± 0.0118, a 5.15 $\sigma$ effect with a probability of 2.5x10$^{-7}$ for occurring by chance. This result used data selection cuts very similar to those in Ref. 3 and contained about 7 times as many galaxies; the results of the two studies are in excellent agreement. The axis lies near $(\alpha, \delta)$ =(216°, 25°), or $(l, b)$ =(32°,69°) in Galactic coordinates with the sense of the axis defined to be along the direction of the $L$ (↺) excess. The uncertainty in space angle of the axis is ~17°. The cosine dependence of the asymmetries is compatible with a pure dipole with a significance ~3.7$\sigma$, the monopole (bias) term is consistent with 0. This result is consistent with the spin asymmetry of 0.047±0.029 with a probability of 11% found in an analysis of the Iye and Sugai spin catalog in the opposite hemisphere.

This axis is 21° away from the north pole of our galaxy (NGP) at $b$=90°. Our galaxy has its spin vector nearly aligned with the preferred axis of spiral galaxies, corresponding to a probabil-ity $(1-\cos 21°)/2 = 0.3$ for this to occur by chance of 3.3%.[d] Inspection of Fig. 1 shows there is no obvious redshift dependence of the asymmetry out to $z \approx 0.085$, which is well beyond the scale of superclustering. Extension of this study to larger redshifts will be difficult due to the problem of reliably recognizing the handedness of faint galaxies.

---

[c] See Table 2 of Lintott et al. The uncertainties in the asymmetry are calculated as $\sigma(A) = 1/\sqrt{R+L}$.

There is now a vast literature on possible observations of cosmic anisotropies and explanations thereof. I shall only briefly comment on the observational effects. Many of these revolve around apparent anomalies in the Wilkinson Microwave Anisotropy Probe one-year, three-year[11], and five-year data[12]. For example, analyzing the one-year data, Land and Magueijo[13] find an unlikely alignment of the low $l$ multipoles and a correlation of azimuthal phases between $l = 3$ and $l = 5$ with an apparent axis $(l, b) = (260°, 60°)$, an alignment they refer to as the "axis of evil". Using the three-year WMAP data, Copi et al.[14] find a similar correlation of low $l$ multipoles and a significant lack of correlations for scales >60°. Analyzing the WMAP five-year and three-year data, Bernui[15] finds a significant asymmetry in large-angle correlations between the north and south Galactic hemispheres at >90% confidence level, depending on the map used. Su and Chu[16], in a recent reanalysis of the WMAP data, find a general alignment of the directions for $l = 2$ to 10 modes to within about $1/4^{th}$ of the northern Galactic hemisphere at latitudes between about 45° and 85°. There is also a considerable literature on a "Virgo alignment" toward $(l, b) \sim (281°, 75°)$ of axes related to the CMB as well as other anisotropic effects in polarizations of radio galaxies and the optical frequency polarization correlation of quasars.[17] All of these effects seem to lie at high Galactic latitudes with a wide spread in longitude. It is also notable that the axis I find at $(l, b) = (32°, 69°)$ lies almost directly opposite the direction of the cold spot in the CMB found by Vielva et al.[18] at $(l, b) = (209°, -57°)$ with an angular size ~10°.

I am extremely grateful to the SDSS group whose efforts and dedication made this work possible. I am indebted to Dr. Masanori Iye for providing the spin catalog of southern galaxies. Undergraduates E. Mallen, A. Bomers, J. Middleton, and M. Pearce made important contributions in their diligent work as scanners. B. McCorkle did the JAVA/HTML programming.

---

[d] The combined probability of the spin asymmetry found here, that for the Iye and Sugai catalog, and the near alignment of the NGP is $\sim 10^{-9}$.